\documentclass[twocolumn,floatfix,prl,showpacs,superscriptaddress]{revtex4-2}

\usepackage{graphicx}
\usepackage{amsmath}
\usepackage{xcolor}

\usepackage{braket}

\begin{document}
    \title{Dimensional reduction and incommensurate  dynamic correlations  in the $S=\frac{1}{2}$ triangular-lattice  antiferromagnet Ca$_3$ReO$_5$Cl$_2$}

\author{S.~A.~Zvyagin}
\thanks{Corresponding author: s.zvyagin@hzdr.de}
\affiliation{Dresden High Magnetic Field Laboratory (HLD-EMFL) and
W\"urzburg-Dresden Cluster of Excellence ct.qmat, Helmholtz-Zentrum
Dresden-Rossendorf, 01328 Dresden, Germany}

\author{A.~N.~Ponomaryov}
\altaffiliation[Present Address: ]
{Institute of Radiation Physics,  Helmholtz-Zentrum
Dresden-Rossendorf, 01328 Dresden, Germany.}
\affiliation{Dresden High Magnetic Field Laboratory (HLD-EMFL) and
W\"urzburg-Dresden Cluster of Excellence ct.qmat, Helmholtz-Zentrum
Dresden-Rossendorf, 01328 Dresden, Germany}

\author{J.~Wosnitza}
\affiliation{Dresden High Magnetic Field Laboratory (HLD-EMFL) and
W\"urzburg-Dresden Cluster of Excellence ct.qmat, Helmholtz-Zentrum
Dresden-Rossendorf, 01328 Dresden, Germany}
\affiliation{Institut f\"{u}r Festk\"{o}rper- und Materialphysik,
TU Dresden, 01062 Dresden, Germany}

\author{D.~Hirai}
\affiliation{Institute for Solid State Physics, University of Tokyo, Kashiwa, Chiba 277-8581, Japan}
\author{Z.~Hiroi}
\affiliation{Institute for Solid State Physics, University of Tokyo, Kashiwa, Chiba 277-8581, Japan}
\author{M.~Gen}
\affiliation{Institute for Solid State Physics, University of Tokyo, Kashiwa, Chiba 277-8581, Japan}
\author{Y.~Kohama}
\affiliation{Institute for Solid State Physics, University of Tokyo, Kashiwa, Chiba 277-8581, Japan}
\author{A.~Matsuo}
\affiliation{Institute for Solid State Physics, University of Tokyo, Kashiwa, Chiba 277-8581, Japan}
\author{Y.~H.  Matsuda}
\affiliation{Institute for Solid State Physics, University of Tokyo, Kashiwa, Chiba 277-8581, Japan}
\author{K.~Kindo}
\affiliation{Institute for Solid State Physics, University of Tokyo, Kashiwa, Chiba 277-8581, Japan}

\date{\today}

\begin{abstract}

The   observation of spinon excitations in the $S=\frac{1}{2}$  triangular antiferromagnet  Ca$_3$ReO$_5$Cl$_2$ reveals a quasi-one-dimensional (1D) nature of  magnetic correlations,  in spite of the  nominally 2D magnetic structure.  This phenomenon is  known as frustration-induced dimensional reduction.   Here, we present   high-field electron spin resonance  spectroscopy and  magnetization studies of  Ca$_3$ReO$_5$Cl$_2$,  allowing us not only to refine  spin-Hamiltonian parameters, but also to investigate peculiarities of its   low-energy spin dynamics. We argue  that the  presence of the uniform  Dzyaloshinskii-Moriya interaction (DMI) shifts  the spinon continuum in momentum space and, as a result,  
opens   a zero-field gap at the $\Gamma$ point.  We observed this gap directly. The shift is found to be consistent with the structural modulation  in the  ordered state, suggesting  this material as a perfect model triangular-lattice system, where a pure DMI-spiral ground state can be realized.

\end{abstract}

%\pacs{75.10.Jm, 75.50.Ee, 76.30.-v, 75.30.Et}
\maketitle

$S=\frac{1}{2}$ antiferromagnetic  systems with triangular  structures are in the focus of modern quantum physics, 
 in particular,  in connection with Anderson's  idea of  “resonating valence bond” (RVB) states  in frustrated spin systems 
\cite{Anderson}. He  proposed that the corresponding  ground state can be  a two-dimensional (2D) fluid  of
resonating spin-singlet pairs, with the elementary excitation spectrum formed by fractionalized mobile quasiparticles, spinons.  Such excitations were observed in the spatially anisotropic  triangular-lattice antiferromagnet (AF)  Cs$_2$CuCl$_4$ \cite{Coldea_2D_SL,Coldea_Cont}, suggesting that the spin-liquid scenario is indeed  realized in this material.  On the other hand, 
more recent   analysis of the  inelastic neutron scattering data \cite{Kohno} unveiled  a  quasi-1D nature of magnetic correlations  in Cs$_2$CuCl$_4$, in spite of its  nominally 2D magnetic structure.  This  phenomenon is  known as  frustration-induced dimensional reduction \cite{Balents}. Later, electron spin resonance (ESR) spectroscopy studies of Cs$_2$CuCl$_4$ in the magnetically disordered state revealed the  presence of an energy gap \cite{Povarov}, corresponding to a shift of the spinon continuum in momentum space, as predicted for an $S=\frac{1}{2}$ isotropic Heisenberg AF chain  with  uniform antisymmetric exchange interaction  (also known as uniform  Dzyaloshinskii-Moriya interaction; DMI) \cite{Bocquet}.     Recently, the quasi-1D nature of  spin correlations in Cs$_2$CuCl$_4$ has been independently  confirmed by thermal-transport measurements \cite{Schulze}. It was shown  that the  uniform DMI remains   playing  an important  role also  below $T_N$,  favoring a non-collinear  helical  spin structure   \cite{Coldea_CCC}.

\begin{figure} [!h]

\begin{center}
%\vspace{0.5cm}
\hspace*{-1.2cm}
\includegraphics[width=0.6\textwidth]{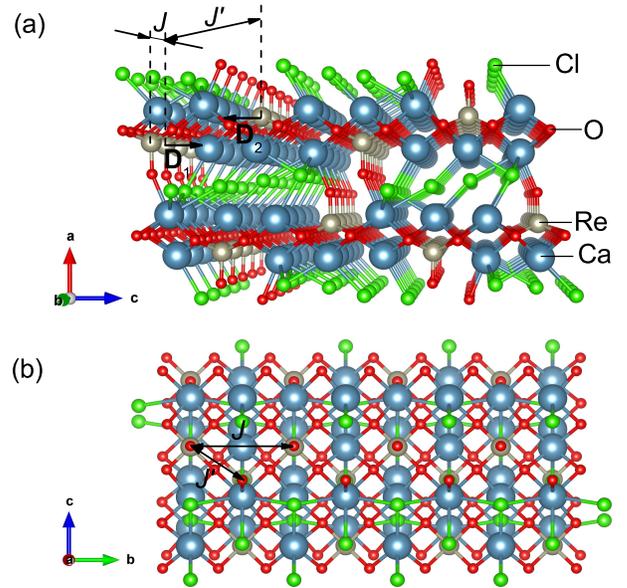}
%\vspace{-1}
\caption{\label{fig:FIG_STR} Crystal structure of CROC.  (a) Perspective view  along the Re$^{6+}$  chain direction ($b$ axis).   Ca atoms are shown in blue, Re in gray, O in  red,  and Cl in green.  $J$ and $J'$ are intra- and interchain exchange interactions, respectively.  The DMI vectors from adjacent chains  (\textbf{D}$_1$ and \textbf{D}$_2$) are  schematically shown by arrows (see the text for details). (b) Parallel view of the crystal structure of CROC  along the $a$ axis.}

\end{center}
\end{figure} 
%Dzyaloshinskii–Moriya interaction (DMI) is an asymmetric magnetic exchange between magnetic moments 

Such DMI-induced incommensurate magnetic structures have recently attracted great attention, hosting a number of intriguing phenomena (e.g.,   magnetoelectric effects in multiferroics \cite{Mostovoy,SWC} and   magnetic skyrmions \cite{Tokura}), which have a direct relevance to various  potential technological applications, including  sensors, magnetic-memory devices, etc.   Since competing  nearest- and next-nearest-neighbor  exchange interactions  can be  another source of the magnetic  incommensurability \cite{Bur,White}, 
 it is important to distinguish between these two principally different  mechanisms, which  can coexist  in real materials \cite{Room}.   Searching for new model materials with  non-collinear incommensurate structures, where  spin correlations are  determined solely by  DMI  (i.e., without or with a minimal   admixture of  competing exchange interactions) remains a very important  task, both from the scientific as well as application perspective.

The recently synthesized compound Ca$_3$ReO$_5$Cl$_2$ (CROC hereafter) is known for its  unusually pronounced  pleochroism \cite{Ple}, exhibiting different colors depending on the viewing direction. The compound crystallizes in an orthorhombic $Pmna$ ($Z=4$) structure   with   magnetic Re$^{6+}$ ions arranged in triangular-lattice structures in the $bc$ plane (Fig.~\ref{fig:FIG_STR}) \cite{Str,Hir_2}. Each Re$^{6+}$ ion is surrounded by five oxide ions,  forming a ReO$_5$ square-pyramidal unit (Fig.~\ref{fig:FIG_STR}~(a)).  The crystal field from the neighboring 
oxide and chloride ions lifts the degeneracy of the Re 5$d^1$ levels, stabilizing  the effective $S=\frac{1}{2}$ $d_{xy}$ orbital ground state with the $d_{xy}$ orbitals confined in the $bc$ plane.   The   orbitals  overlap  with each other, resulting in a large direct exchange interaction $J$  along the $b$ axis (this axis is  the chain direction).   A small orbital  overlap between adjacent chains results in the weaker  interchain  exchange interaction  $J'$.  The Ca$_3$ReO$_5$ layers are  well  separated from each other by Cl layers. First-principle density functional theory (DFT)  calculations revealed that the interplane exchange interaction $J''$ is more than three orders of magnitude smaller than the leading interaction  $J$ \cite{Str}, allowing to map CROC to a quasi-2D triangular-lattice AF model. ReO$_5$ pyramidal  units from the same chain point in the same direction, while  units from 
adjacent chains point in opposite direction. As a result, there is no inversion center between neighboring Re atoms along the chains, allowing uniform DMI  on the $b$-bond.

Recent inelastic neutron-scattering studies of this material revealed the  presence of a two-spinon continuum \cite{Nawa},  confirming  the dimensional reduction in this   $S=\frac{1}{2}$ AF with triangular-lattice structure.   Similar to Cs$_2$CuCl$_4$, the quasi-1D character of the magnetic correlations  is  evident   from  a particular pronounced   dispersion of  magnetic excitations along the $b$ axis and a  distinct asymmetry of the neutron-scattering intensities  in  momentum space (a signature of bound spinon excitations, triplons). The quasi-1D nature of magnetic excitations in CROC was confirmed  by  means of Raman scattering   \cite{Choi}. Neutron-diffraction measurements  revealed an incommensurate magnetic structure in CROC below $T_N=1.13$ K,   with the ordering wavevector  $\textbf{\textit{q}}$ = (0, 0.465, 0) \cite{Nawa}.

Compared to Cs$_2$CuCl$_4$ with 
$J/k_B=4.7$~K and $J'/J\simeq 0.30$  \cite{CCC_NS, CCC_ESR},  CROC is characterized by about one order of magnitude  larger scale of  exchange  interactions. Fit of the magnetic susceptibility using an  anisotropic triangular-lattice  model  unveiled   $J/k_B=40.6$ K and $J'/J=0.32$, while a Bonner-Fisher fit  for the $S=\frac{1}{2}$ Heisenberg AF chain model,   combined with results of DFT calculations, suggested   $J/k_B=41.3$ K and $J'/J=0.295$ \cite{Str}.  On the other hand, recent inelastic neutron-scattering measurements revealed    $J/k_B=41.7$~K and $J'/J=0.15(5)$ \cite{Nawa}.   Thus, although the intrachain exchange parameters are in good agreement with each other, an independent and, preferably, more  accurate estimation of $J'$ in CROC  remains a challenging open problem. Here, we present  high-field ESR spectroscopy and  magnetization  studies of CROC, allowing us not only to refine its spin-Hamiltonian parameters, but also to investigate the  magnetic properties and peculiarities of spin dynamics  in a broad range of frequencies and magnetic fields, relevant to the energy scale of  magnetic interactions in this new  frustrated spin system.

\hfill \break

\textbf{Results}

\begin{figure} [!ht]

\begin{center}
%\vspace{0cm}
\hspace*{-0.5cm}
\includegraphics[width=0.6\textwidth]{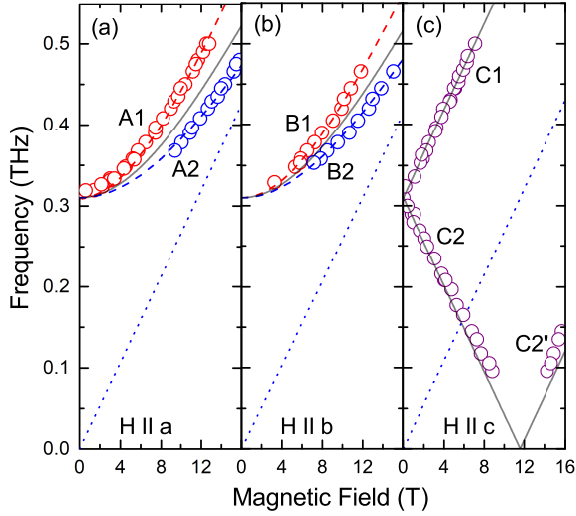}
%\vspace{2cm}
\caption{\label{fig:FIG_FFD} Frequency-field diagrams of magnetic excitations in CROC. Fields are  applied along the $a$ (a),  $b$ (b), and $c$ (c) axis ($T=2$ K). The solid lines are results of calculations using Eqs.  (\ref{ESR2}) and (\ref{ESR1}). The dotted lines correspond to the paramagnetic $g$-factors, 1.88,  1.85, and 1.92 for field applied along the $a$, $b$, and $c$ axis, respectively \protect\cite{Nawa}. The dashed lines are guides for the eye.}
\end{center}
\end{figure}

\begin{figure} [!h]

\begin{center}
%\vspace{0cm}
\hspace*{-0.6cm}
\includegraphics[width=0.6\textwidth]{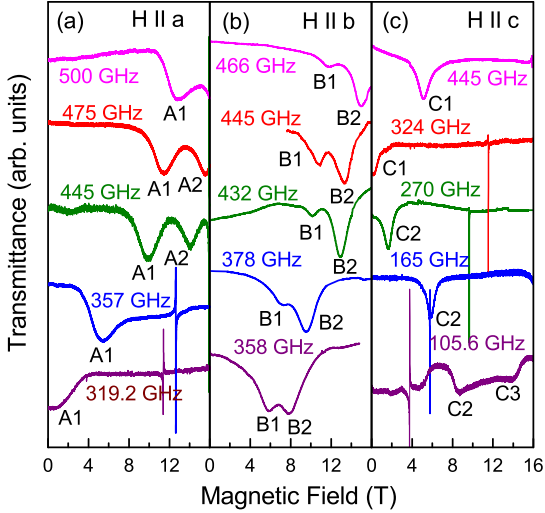}
%\vspace{2cm}
\caption{\label{fig:FIG_SP} Exemplary  ESR spectra. Fields are applied along the $a$ (a), $b$ (b), and $c$ (c) axis ($T=2$ K).  Narrow lines with $g=2$ correspond to  DPPH, used as a marker.}
\end{center}
\end{figure}

\begin{figure} [!h]

\begin{center}

\hspace*{-0.8cm}
\includegraphics[width=0.55\textwidth]{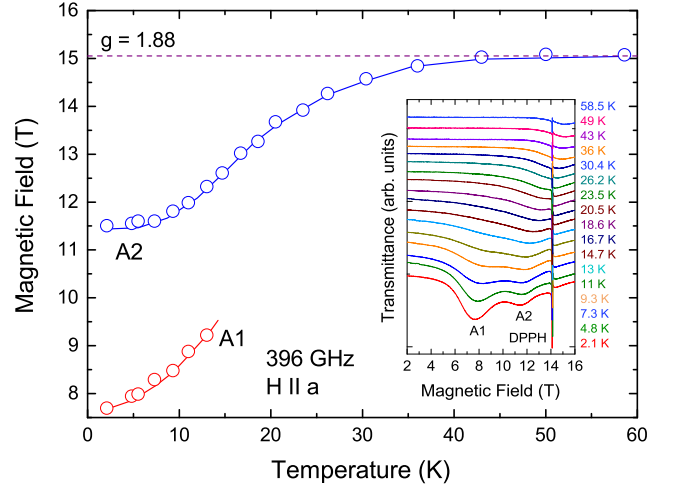}
%\vspace{-0.45cm}
\caption{\label{fig:FIG_TD} (a) Temperature dependences of ESR fields in CROC.  The data are taken at a frequency of  396 GHz with magnetic field applied along the $a$ axis.  The dashed line corresponds to the paramagnetic $g=1.88$.  The solids  lines are guides for the eye.  The inset shows  corresponding spectra at different temperatures (narrow lines with $g=2$ correspond to  DPPH, used as a marker).}
\end{center}
\end{figure}

\textbf{High-field ESR.} 
In Fig.~\ref{fig:FIG_FFD}, we show the frequency-field diagrams of the  ESR excitations measured at a temperature of 2 K with magnetic field applied  along the $a$, $b$, and $c$ axes.  We estimate the accuracy of sample orientation to be better than $\pm 5$ degree (see Supplementary Information). We observed two  ESR modes for  fields applied along the $a$ and $b$ axes (A1, A2 and B1, B2, respectively), while we detected  three modes (C1, C2, C2$'$)   for  the field  along the $c$ axis. The extrapolation of the frequency-field dependences  to zero field reveals a gap of  $\Delta = 310(5)$ GHz. In Fig.~\ref{fig:FIG_SP}, 
we show examples of low-temperature ESR spectra. We measured also  the temperature dependences of the resonance fields for the modes A1 and A2 at a frequency of  396 GHz  (Fig.~\ref{fig:FIG_TD}).

\begin{figure} [!h]

\begin{center}
%\vspace{0cm}
\hspace*{-0.8cm}
\includegraphics[width=0.55\textwidth]{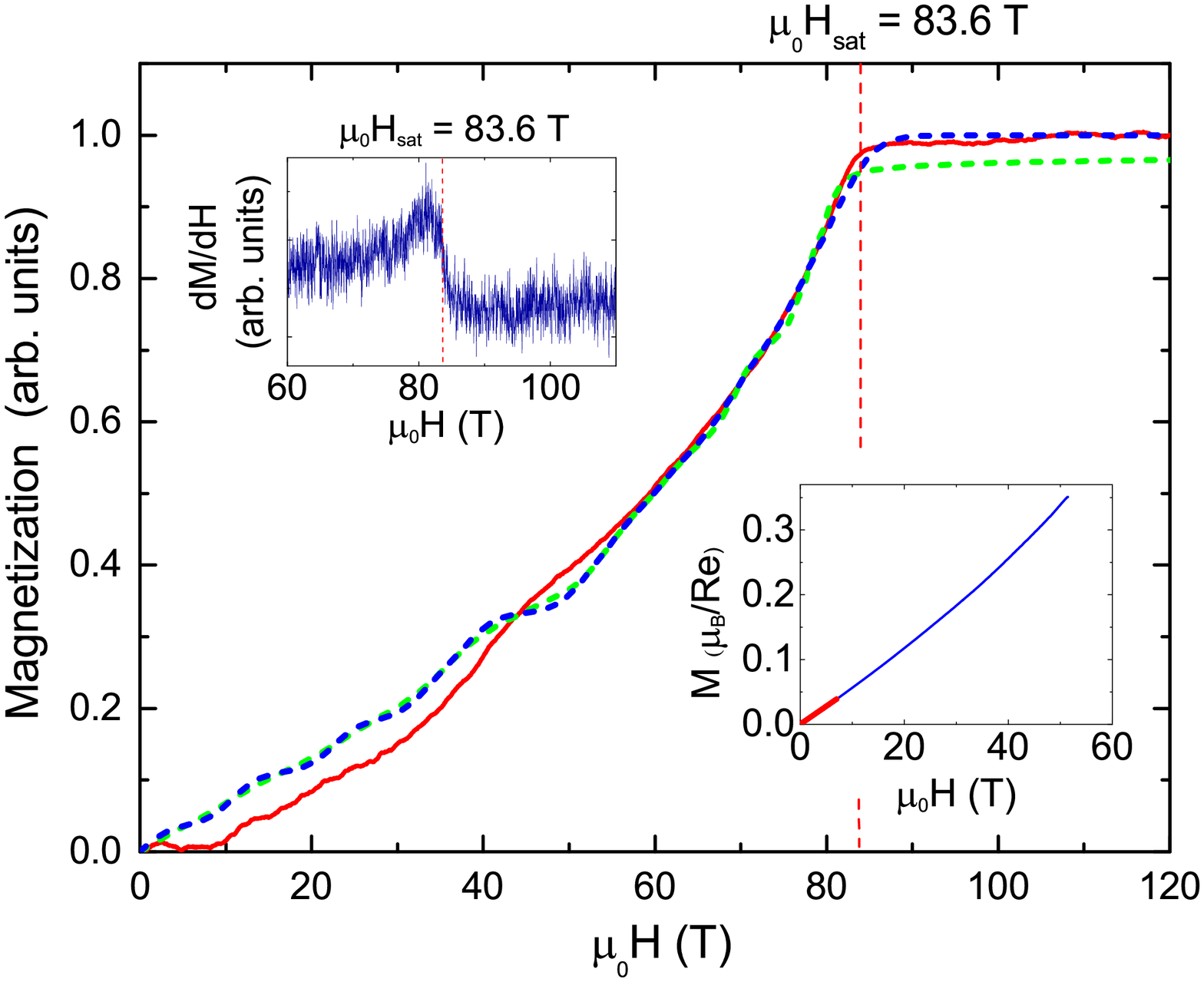}
%\vspace{2cm}
%\vspace{-0.5cm}
\caption{\label{fig:FIG_M} Magnetization of CROC. Main panel: magnetization  of a  CROC powder  sample  in magnetic fields up to 120 T,  
obtained using a pulsed single-turn magnet (red line; the initial   temperature is  5 K) together with results of OFTLM calculations for a triangular-lattice AF with  $J/k_B=41.7$~K,    $J'/J=0.25$ in isothermal (blue  dashed line) and adiabatic (green dashed line) approximations \protect\cite{Morita}. The left inset shows the derivative of the as-measured  magnetization $M$; the saturation field $\mu_0 H_{sat}=83.6 $ T is determined as shown.   The right  inset shows the magnetization of a  CROC powder  sample in magnetic fields up to 51 T measured using a non-destructive magnet (the initial   temperature is 1.3 K).  Magnetization of a powder sample at a temperature of 2 K in DC  fields up to 7 T is shown in red.  }
\end{center}
\end{figure}

\textbf{High-field magnetization.}  In Fig.~\ref{fig:FIG_M}, we show the  magnetization  of a  CROC powder  sample  in magnetic fields up to 120 T,   obtained using a pulsed single-turn magnet (red solid line). The left inset shows the derivative  of the as-measured high-field magnetization, revealing clearly the saturation  field of $\mu_0 H_{sat}=83.6$ T. In addition, we performed magnetization the measurements of a CROC powder  sample   in magnetic fields up to 51 T using a nondestructive  magnet  (blue line, right inset in Fig.~\ref{fig:FIG_M}). The employment of this magnet  allowed us to achieve much  better signal-to-noise ratio and to  identify   low-field features of the 120 T magnetization as artifacts. In addition, we measured the  magnetization of a powder sample at a temperature of 2 K in DC  fields up to 7 T (red line). The experimental data were compared with results of calculations obtained   
using the orthogonalized finite-temperature Lanczos method (OFTLM)  for a  triangular-lattice AF \cite{Morita}.

\hfill \break

\textbf{Discussion}

The  spin dynamics in  an $S=\frac{1}{2}$ Heisenberg AF chain with uniform nearest-neighbor
exchange coupling $J$  is determined by a gapless two-particle continuum of fractional 
$S=\frac{1}{2}$ excitations,  spinons.  The energy of the  lower boundaries of the continuum is  given  by the des Cloizeaux-Pearson relation \cite{dCP,dCP_1}
\begin{eqnarray}
\label{E1}
\varepsilon_l (q)= \frac{\pi}{2} J | \sin (q)|,  
\end{eqnarray}
and the upper boundaries are described  by the formula  \cite{Yamada}
\begin{eqnarray}
\label{E2}
\varepsilon_u (q)= \pi  J | \sin (q/2)|,  
\end{eqnarray}
where $q$ is  the wavevector along the chain direction. 

The  presence of  uniform DMI in an $S=\frac{1}{2}$ Heisenberg AF chain  can significantly modify the excitation spectrum, shifting the spinon continuum in  momentum space  \cite{Bocquet}. As result,   an energy  gap at the Brillouin-zone center opens. Such a gap was observed by means of ESR in the   ``dimensionally reduced'' triangular-lattice AF  Cs$_2$CuCl$_4$  \cite{Povarov},    quasi-1D Heisenberg AFs  Na$_2$CuSO$_4$Cl$_2$ \cite{Fujihala},  K$_2$CuSO$_4$Br$_2$ \cite{Smirnov}, and K$_2$CuSO$_4$Br$_2$ \cite{Soldatov}.   The   employment of ESR techniques  allows one not only to directly measure  the uniform DMI, but also  to  experimentally investigate the interaction between fractionalized spinon excitations, including   backscattering processes \cite{Povarov_PRL,Wang}.

The theory of ESR in an $S=\frac{1}{2}$ Heisenberg AF chain with uniform DMI \cite{Povarov,Smirnov,Wang}  predicts the presence of two or one ESR modes, depending on field direction. The excitation frequency-field diagram for $\textbf{H} \parallel \textbf{D}$ is  given by 
\begin{eqnarray}
\label{ESR2} h\nu_{\pm} =  \displaystyle\left\lvert g_{\parallel}\mu_BH\pm  \frac{\pi D}{2} \right\rvert 
\end{eqnarray} 
(thereby, $D$ is a parameter describing the DMI strength), while for  $\textbf{H }\perp \textbf{D}$  
\begin{eqnarray}
\label{ESR1} h\nu =   \sqrt{(g_{\bot }\mu_BH)^{2} + \left(\frac{\pi D}{2}\right)^2}.
\end{eqnarray}

As mentioned above,   due to the absence of an  inversion center between adjacent Re$^{6+}$ ions along the chains, DMI  in CROC is allowed along  the $b$ direction  (Fig.~\ref{fig:FIG_STR}). There is a mirror plane perpendicular to the  chains with a bisecting point located in the middle of the section connecting  two  neighboring Re$^{6+}$ ions.  Therefore, in accordance with Moriya's rules \cite{Mor}, the  DM vector is   in the $ac$ plane (Fig.~\ref{fig:FIG_STR}). It is important to mention that the  ReO$_5$ pyramidal units are not connected with each other by sharing common oxygen atoms. Instead,  Re$^{6+}$ ions along the  chains are linked by   a more complex  superexchange path, formed by four  basal oxygen and two calcium atoms. The path has  a mirror plane, which includes  one oxygen ion
is in the pyramid apex and two  neighboring rhenium  ions.  Because of that, the  DM vector is expected to be perpendicular to the mirror plane \cite{Mor}, i.e.,  $\textbf{D} \parallel \textbf{c}$. For  $\textbf{H} \parallel \textbf{D}$, the theory \cite{Povarov,Smirnov,Wang} predicts  ESR modes as described by Eq.  (\ref{ESR2}). These modes were indeed  observed in our experiments with $\textbf{H} \parallel \textbf{c}$ (Fig.~\ref{fig:FIG_FFD}~(c)).

For the relevant  magnetic-field range with the field applied perpendicular to \textbf{D}, the theory predicts the presence of one ESR mode (Fig.~\ref{fig:FIG_FFD}~(a,b), solid lines). On the other hand, our analysis of the ESR angular dependence revealed that even a small  (a few degree)  field tilting from the $b$ towards the $c$ direction  may result in an  ESR line splitting, as observed by us for all frequencies above 350 GHz (as an example, ESR angular dependences  at 370 GHz with  magnetic field in  the $bc$ plane  are presented  in Supplementary Information). Most importantly, the extrapolations of the ESR frequency-field dependences to zero field  suggest   the presence of a zero-field gap, $\Delta = 310$ GHz. Employing  Eqs.  (\ref{ESR2}) and (\ref{ESR1}) we can estimate DMI, yielding  $D/k_B =9.47$ K.

The temperature dependences of the  resonance fields  for  the modes A1 and A2 (Fig.~\ref{fig:FIG_TD}) revealed   that with increasing temperature  these modes come so close to each other,   that  above 13 K one  can  resolve only one line.  On the other hand, the ESR absorption   shifts to higher fields, reaching the paramagnetic value $g=1.88$ at  about 40 K (dashed line  in Fig.~\ref{fig:FIG_TD}). This temperature   corresponds to the energy of the intrachain exchange interaction $J$; above this temperature  thermal fluctuations become dominant, significantly suppressing spin correlations along the chain direction.  Investigating  temperature and angular dependences   of  ESR parameters  (which in CROC are dominantly  determined by the  exchange interaction $J$ and  uniform DMI $D$) remains a topic  of  further experimental and  theoretical investigations.

\begin{figure} [!h]

\begin{center}

\hspace*{-0.5cm}
\includegraphics[width=0.55\textwidth]{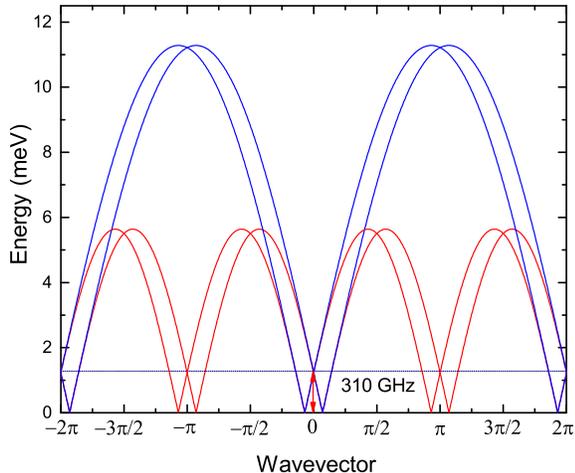}
%\vspace{-0.5cm}
\caption{\label{fig:FIG_DISP} Two-spinon continuum  calculated  for an $S=\frac{1}{2}$ Heisenberg AF chain with $J/k_B=41.7$~K   and  the uniform DMI  $D/k_B=9.47$ K. The upper and lower boundary of the two-spinon continuum are denoted by the blue and red lines, respectively. The gap observed at the $\Gamma$ point,  $\Delta=310$ GHz,    is shown by the arrow. 
 }
\end{center}
\end{figure}

Knowing the  saturation field  and intrachain exchange coupling, we  now can  determine the interchain exchange interaction $J'$ from  the expression

\begin{equation}
g\mu_B H_{sat} = 2J(1+J'/2J)^2,
\label{ccc_sat}
\end{equation}
obtained from  the exchange model of a triangular-lattice AF \cite{CCC_ESR}. Using the averaged value 
$\braket{g}=1.883$   for the powder sample, $J/k_B=41.7$~K from inelastic neutron scattering   \cite{Nawa},  and  $\mu H_{sat}=83.6 $ T (Fig.~\ref{fig:FIG_M}), 
we obtain   $J'/k_B=10.5$ K  and $J'/J=0.25$.

In Fig.~\ref{fig:FIG_M}, together with experimental magnetization data (red line) we show also OFTLM calculation results  for a  triangular-lattice AF   with  $J/k_B=41.7$~K and $J'/J=0.25$ \cite{Morita}.  The data   for  an isothermal ($k_BT/J = 0.05$; $N = 36$ sites) approximation   are shown by the blue dashed line. It is important to mention that, due to the relatively small pulse duration,  magnetization processes in pulsed fields in the megagauss range  are  not isothermal, but rather adiabatic. The  adiabatic magnetization  for $J/k_B=41.7$~K, $J'/J=0.25$, and $S_m/N=0.075$ (where $S_m$ is the magnetic anisotropy and $N = 36$ sites) is  shown in Fig.~\ref{fig:FIG_M} by  the green dashed line.  Very good agreement between the experimental data and the calculation is  achieved. For the adiabatic process at  a given initial temperature ($\sim 0.1 J/k_B$) the theory  suggests  that the ground state at  $H_{sat}$ is only partially spin polarized  (green dash line in Fig.~~\ref{fig:FIG_M}.

In Fig.~\ref{fig:FIG_DISP}, we schematically  show  a two-spinon continuum  for  $S=\frac{1}{2}$ Heisenberg AF chain system  with $J/k_B=41.7$~K and the $\Gamma$-point gap $\Delta=310$ GHz, as revealed  by ESR. For these values, the shift of the soft mode in the magnetically disordered phase  (determined by the  uniform DMI, as discussed above) corresponds to the incommensurate wavevector   $\textbf{\textit{q}}=0.464$, which   is remarkably consistent  with  the  ordering wavevector  $\textbf{\textit{q}}$ = (0, 0.465, 0) below $T_N$ \cite{Nawa}. This should  be regarded as  an  important prerequisite for the realization of  a pure DMI-spiral magnetically ordered state in this material at low temperatures.

The situation is very different from  the case of Cs$_2$CuCl$_4$, where the incommensurate  ordered structure appears to be a combined effect 
of the frustration,  induced by exchange interaction between the spins along chains \cite{Coldea_CCC}, and DMI. Assuming   $J/k_B=4.7$~K \cite{CCC_NS, CCC_ESR} and $\Delta = 14$ GHz  \cite{Povarov}, we obtain  for the soft mode in the magnetically disordered phase   $\textbf{\textit{q}}=0.485$. This value is somewhat  different from the $b$ component of the ordering wavevector $\textbf{\textit{q}}$ = (0, 0.472, 0) \cite{Coldea_CCC}, which results  in a twice larger gap at the $\Gamma$ point \cite{Smirn}.  The difference between Cs$_2$CuCl$_4$ and CROC can be understood, taking into account  that CROC is characterized by an about  five times  larger frustration factor $f=\lvert\Theta_{cw}\rvert/T_N$  (33.5 vs 6.5  for CROC and Cs$_2$CuCl$_4$, respectively \cite{Str,Tokiwa};  where $\Theta_{cw}$ is the Curie-Weiss temperature).  The larger frustration in CROC    leads to  more effective isolation of neighboring chains from each  other,  with the uniform DMI playing the key role above and  below $T_N$.

In summary, we performed high-field ESR spectroscopy and magnetization studies of  CROC allowing us to characterize this material as 
a spatially anisotropic triangular-lattice Heisenberg  AF with  $J'/J=0.25$, frustration-induced  dimensional reduction, and the incommensurate spin dynamics.    Our findings illuminate    the important role  of the uniform DMI in this material,  affecting  the spin dynamics in the magnetically disordered state and determining  peculiarities of its  magnetic structure in the magnetically ordered phase. Based on our observations, a pure DMI-spiral state can be realized in CROC,  making  this material  an attractive toy model to explore details of  the dimensional reduction and other  effects of the geometrical frustration in low-D spin systems with competing interactions.

\hfill \break

\textbf{Methods}

\textbf{Single-crystal growth}.  Single crystals of CROC were grown by a flux method \cite{Hir_2}.  In an argon-filled glovebox,
CaO, ReO$_3$, and CaCl$_2$ were mixed in an agate mortar in a 4.1:1:17 molar ratio,
then placed in a gold tube and sealed in an evacuated quartz ampule.
The ampule was heated to  1000$^{\circ}$~C and kept  for 24 hours before being cooled down to 800$^{\circ}$~C at a rate of 1$^{\circ}$~C per hour.
Single crystals  were obtained after washing away excess CaCl$_2$ flux with distilled water. Single crystals with typical sizes of ca 4x4x1 mm$^3$ were used in ESR experiments. 
The crystal's cleavage plane is perpendicular to the $a$ axis. It is important to mention that the single crystals gradually decompose in air due to moisture.

\textbf{High-field ESR}.  High-field ESR measurements of CROC were performed at the Dresden High Magnetic Field Laboratory (HLD) using  a transmission-type ESR spectrometer (similar to that described in Ref. \cite{Zvyagin_INSTR})  with oversized waveguides and a 16 T superconducting magnet.  A set of backward-wave oscillators, Gunn diodes, and VDI microwave sources (Virginia Diodes Inc, USA) was used,
allowing us to study  magnetic excitations 
in a  broad quasicontinuously covered frequency range,  from 100 
to 500 GHz. The experiments were performed 
in the Faraday and Voigt    configurations  
with magnetic fields $H \parallel a$ and $H \parallel b,c$, respectively.   The stable free-radical 2,2-diphenyl-1-picrylhydrazyl (DPPH)  was used as a frequency-field marker.

\textbf{High-field magnetization}.     The  megagauss-field   facility at the ISSP, University of Tokyo, Japan, equipped with a 160 $\mu$F/50 kV capacitor bank,  was used to generate  ultrahigh  magnetic fields  \cite{Takeyama, Miura}.   To measure magnetization, we employed  two types of magnet. For 120 T experiments we used    single-turn coil magnets  in   horizontal geometry, with the pulse duration 8 $\mu$s. The induction method was used  to detect the d$M$/d$H$ signal using  coaxial-type pickup coils.   The field was measured by a calibrated pickup coil located near the sample. The recorded pickup voltage was numerically integrated to obtain  field values.  In spite of a careful compensation,  the  pulsed-field magnetization  exhibited  a tiny linear background, which we subtracted from the experimental data. Magnetization up to 51 T was measured using a non-destructive  magnet with a pulse duration  of 4 ms. More detailed information on the experimental procedure can be found in Ref. \cite{Gen}.  The magnetization was calibrated to absolute values using  the magnetization data collected  at 2 K in DC  fields up to 7 T   using a  SQUID magnetometer (Quantum Design, USA).

\hfill \break

\textbf{Data availability}

The data that support the findings of this study are available from the corresponding 
author upon reasonable request.

 \hfill \break

\textbf{Supplementary Information}

\begin{figure} [!h]

\begin{center}

\hspace*{-0.55cm}
\includegraphics[width=0.55\textwidth]{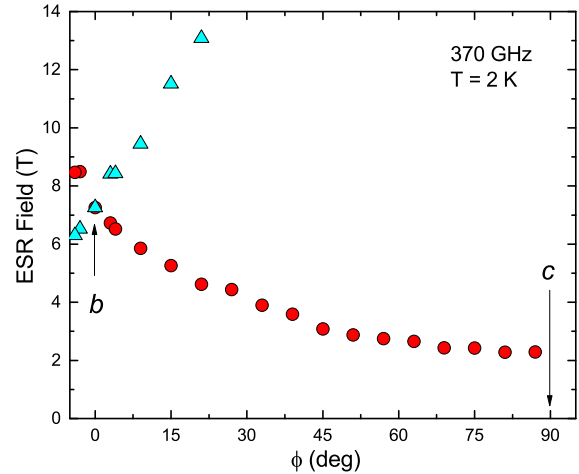}
%\vspace{-0.5cm}
\caption{\label{fig:FIG_AD} ESR  angular dependence  for magnetic field in  the $bc$ plane at a frequency of 370 GHz ($T=2$ K). Two observed modes are denoted by different colors.
 }
\end{center}
\end{figure}

For high-field  electron-spin resonance (ESR) experiments
 single crystals with typical sizes of ca 4x4x1 mm$^3$ (along the $a$, $b$, and $c$ axis, respectively) were used. In  Fig.~\ref{fig:FIG_AD}  we show 
 the  angular dependence of ESR absorptions  for magnetic field applied in  the $bc$ plane at a frequency of 370 GHz. The measurement  were performed at a temperature of 2 K, clear indicating  a large anisotropy of ESR absorption, in particular  in the vicinity of the $b$ axis. As revealed,  even a small  (a few degree)   tilting of magnetic field from the $b$ axis  results in a significant    ESR line splitting. Based on  that, we estimate the  accuracy of sample orientation in our experiments (Fig. 2 - 4)  to be better than $\pm 5$ degree.   The  anisotropy of the observed ESR absorption in the vicinity of the $c$ axis is not as pronounced. The mode C1 was observed with magnetic field of about 2 T applied along the $c$ direction, while much higher field is needed for the observation of the mode C2$'$ at this frequency.

\textbf{Author contributions}

S.Z. conceived, designed and led the project. D.H. and Z.H. grew CROC single crystals. S.Z. and A.P.  performed high-field ESR experiments. 
M.G., Y.K., A.M., Y.M., and K.K.  performed high-field magnetization experiments. J.W.  administered the HLD parts of the project.  All authors discussed the results and commented on the manuscript.

\textbf{Acknowledgment}

This work was supported by the Deutsche Forschungsgemeinschaft, through ZV 6/2-2, the W\"{u}rzburg-Dresden Cluster of Excellence on Complexity and Topology in Quantum Matter - $ct.qmat$ (EXC 2147, project No. 390858490),  and SFB 1143, as well as by HLD at HZDR,  member of the European Magnetic Field Laboratory (EMFL). S.Z. acknowledges the support of the BMBF via DAAD (Project ID. 57457940). This work was partly supported by the Japan Society for the Promotion of Science (JSPS) KAKENHI Grant Numbers JP19H04688 and
JP20H01858.  S.Z.  would like to acknowledge  fruitful  discussions with  A.~I.~Smirnov, K.~Yu.~Povarov, and K.~Nawa. We would like to thank  K.~Morita, who shared with us results of his numerical calculations.

\textbf{Competing Interests}

The authors declare no competing  interests.

\end{document}